\documentclass[namedreferences]{SolarPhysics}
\usepackage[optionalrh]{spr-sola-addons} 
\usepackage{graphicx}
\usepackage{color}           
\usepackage{url}             
\usepackage{epsfig}

\newcommand{\arcsec}{{$^{\prime\prime}$}}

\newcommand{\etal}{{\it et al.}}

\newcommand{\aap}{    {\it Astron. Astrophys.}}

\newcommand{\apj}{    {\it Astrophys. J.}}

\newcommand{\pasj}{   {\it Pub. Astron. Soc. Japan}}

\newcommand{\solphys}{{\it Solar Phys.}}

\newcommand{\ssr}{    {\it Space Sci. Rev.}}

\begin{document}

\begin{article}

\begin{opening}

\title{Radioheliograph Observations of Microwave Bursts with Zebra Structures}

\author{A.T.~\surname{Altyntsev}$^{1}$\sep
        S.V.~\surname{Lesovoi}$^{1}$\sep
        N.S.~\surname{Meshalkina}$^{1}$\sep
        R.A.~\surname{Sych}$^{2,1}$\sep
        Y.~\surname{Yan}$^{2}$
       }
\runningauthor{Altyntsev et al.}
\runningtitle{Zebra structures}

   \institute{$^{1}$Institute of Solar-Terrestrial Physics SB RAS,Lermontov St.\ 126A, Irkutsk 664033, Russia email: \url{altyntsev@iszf.irk.ru} \\
       $^{2}$  Key Laboratory of Solar Activity, National Astronomical Observatories, CAS, Beijing 100012, China e-mail:
       \url{yyh@bao.ac.cn} \\
             }

\begin{abstract}
The so-called zebra structures in radio dynamic spectra,
specifically their frequencies and frequency drifts of emission
stripes, contain information on the plasma parameters in the
coronal part of flare loops. This paper presents observations of
zebra structures in a microwave range. Dynamic spectra were
recorded by Chinese spectro--polarimeters in the frequency band
close to the working frequencies of the Siberian Solar Radio
Telescope. The emission sources are localized in the flare
regions, and we are able to estimate the plasma parameters in the
generation sites using X-ray data. The interpretation of the zebra
structures in terms of the existing theories is discussed. The
conclusion has been arrived that the preferred generation
mechanism of zebra structures in the microwave range is the
conversion of plasma waves to electromagnetic emission on the
double plasma resonance surfaces distributed across a flare loop.

\end{abstract}
\end{opening}


\section{Introduction}
     \label{S-Introduction}

Narrow-band structures in the dynamic spectra of radio bursts have
attracted much attention from solar radio astronomers. Their
coherent emission is generated by a nonthermal electron component,
and its characteristics may be used to determine the plasma
parameters in flare regions close to the site of their initial
energy release. The most remarkable among these structures is the
so-called zebra structure observed in the dynamic spectra as
several, concurrently changing stripes of higher emission
intensity.

The frequencies of the zebra emission must be above the cutoff
frequency that is the Langmuir frequency in the source. Structures
in meter-- and decimeter--wave ranges are generated in low-density
plasmas in the upper corona at heights over several tens of
thousands of kilometers ({\it e.g.}, see \opencite{Chernov2006a}).
In this case, a large number of stripes drifting to low
frequencies are regularly observed. The sources of coherent
emission at frequencies above 3 GHz should be located in dense
plasma loops in the low corona. Zebra structures can rarely be
seen in this frequency range. Among two hundred events with
sub--second pulses recorded by the Siberian Solar Radio Telescope
(SSRT, \opencite{Grechnev}) and given in the catalogues
(\url{http://badary.iszf.irk.ru/Ftevents.php} and
\url{http://ssrt.iszf.irk.ru/fast}), we found only six flares with
zebra structures observed in the dynamic spectra recorded with
Chinese spectro--polarimeters (\opencite{Fu}; \opencite{Ji}).

The generation of coherent narrow-band bursts in the microwave
range is associated with two-stage plasma mechanisms. At first,
electrostatic plasma oscillations are excited. Then a part of the
oscillation energy is converted into electromagnetic waves at
frequencies close to the Langmuir frequency or to its harmonics.
First of all, the zebra events with stripe frequencies divisible
by the electron gyro-frequency were detected in the microwave
range.

The frequency structure can be explained by the excitation of the
Bernstein modes in compact homogeneous sources
\cite{Altyntsev2005}. The interpretation of events found later
required other models because, as opposed to the Bernstein modes,
the difference between the frequencies of adjacent stripes was not
constant or was too small in comparison to the independent
estimations of the cyclotron frequency. To explain such
structures, \inlinecite{Chernov2006b} suggested, like the case for
low-frequency zebra structures, that the emission of bright
stripes is generated in some compact sources distributed along a
flare loop. Here the emission frequency of a single strip is
determined by the values of plasma density and magnetic field in
its source.

\inlinecite{Chernov2006b} attributed the narrow-band emission to
the excitation of upper hybrid $f_{\rm up}=\sqrt{f_{\rm
p}^2+f_B^2}$ and whistler waves $f_{\rm w}<f_B$ by nonthermal
electrons with a loss-cone velocity distribution. Here $f_{\rm
p}=\sqrt{\frac{ne^2}{\pi{m}}}$ is the Langmuir frequency and
$f_B=\frac{eB}{2\pi{mc}}$ is the electron gyro--frequency. It is
assumed that both types of plasma oscillation are excited
simultaneously by the same population of nonthermal electrons.
Plasma waves are converted into high-frequency electromagnetic
waves when they merge together: $f=f_{\rm up}+f_{\rm w}$. As the
whistler frequency $f_{\rm w}$ is low, the frequency of the
electromagnetic wave $f$ is close to the upper hybrid one.

Note that the closeness of the electromagnetic wave frequency to
the Langmuir frequency results in its rapid damping in the plasma
surrounding the source \cite{Benz} The most controversial point in
the model of \inlinecite{Chernov2006b} is the assumption that the
appearance of a number of quasi-harmonic stripes is due to
discrete distribution of whistler-wave sources along a flare loop.

\inlinecite{Zlotnik2011} pointed out that the intensity of the
electromagnetic emission does not depend considerably on the level
$T_{\rm w}$ of low-frequency whistler turbulence. This conclusion
follows from the Manley-Rowe relation: $T=\frac{fT_{\rm up}T_{\rm
w}}{f_{\rm up}T_{\rm w}+f_{\rm w}T_{\rm up}}$. It is evident that
at $f_{\rm up}\gg{f_{\rm w}}$ the brightness temperature of the
electromagnetic emission is $T \cong T_{\rm up}$. Hence, to
explain the discrete distribution of radio sources, we should
assume a discrete distribution of regions with upper-hybrid
oscillation turbulence. The high-frequency plasma turbulence is
expected to be enhanced in compact regions in which the condition
for double plasma resonance (DPR) $f_{\rm up}=sf_B$ is fulfilled,
where $s$ is an integer (\opencite{Zlotnik2003};
\opencite{Kuznetsov2007b}; \opencite{Kuznetsov2008}).

The observed characteristics of a multiple-strip zebra structure
in the meter-wave range have been interpreted as due to the
emission sources on the DPR surfaces (\opencite{Aurass};
\opencite{Zlotnik2003}). By comparing the frequencies of zebra
stripes and the calculated values of gyro-frequency along the
loop, the authors managed to obtain the plasma density as a
function of height. It proved to be close to the expected
hydrostatic distribution at a reasonable coronal plasma
temperature of about 1 million degrees.

This paper discusses the observations of microwave zebra
structures whose dynamic spectra were recorded by the
spectro-polarimeters of Chinese observatories in a frequency range
of 5.2--7.6 GHz comprising the receiving frequency of SSRT (5.7
GHz). The observational techniques will be described in Section 2.
Using the SSRT interferometer observations, we were able to
localize the microwave sources in the flare area and thus
independently estimate the plasma parameters in the generation
region using X-ray data. Section 3 will describe the observations
of zebra structures. Section 4 is concerned with the applicability
of the DPR mechanism to the interpretation of the generation of
the zebra structure.

\section{Instrumentation}

The techniques for the observation of zebra structures have been
described by \inlinecite{Altyntsev2005}. The dynamic spectra of
zebra structures were observed with the Solar Radio Broadband Fast
Dynamic Spectrometers (5.2--7.6 GHz, both circular polarizations,
frequency and time resolution of 20 MHz and 5 ms, respectively) at
the Huairou Solar Observing Station of the National Astronomical
Observatories of China (NAOC) and the Purple Mountain Observatory
(4.5--7.5 GHz, only intensity, 10~MHz, 5~ms) (\opencite{Fu};
\opencite{Ji}).

The spatial structure of the microwave sources was recorded by the
Siberian Solar Radio Telescope (SSRT) \cite{Grechnev}. Our
investigation of the flare bursts with high time resolution are
based on the data recorded by the EW and NS arrays, which
simultaneously provide one-dimensional images (scans) of the solar
disk every 14 ms. The analysis methods for one-dimensional solar
images have been described by \citeauthor{Altyntsev96}
(\citeyear{Altyntsev96}, \citeyear{Altyntsev2003}) and
\inlinecite{Lesovoi}. The receiver system of SSRT contains a
spectrum analyzer with 120~MHz frequency coverage using an
acousto-optic detector with 250 frequency channels, which
correspond to the knife-edge-shaped fan beams for the NS and EW
arrays. The frequency channel bandwidth is 0.52~MHz. The response
at each frequency corresponds to the emission from a narrow strip
on the solar disk whose position and width depend on the
observation time, array type, and frequency. The signals from all
the channels are recorded simultaneously and generate a
one-dimensional distribution of solar radio brightness. Each of
the right and left components of circular polarization is recorded
successively within the intervals of 7 ms. The minimal width of
the beam of SSRT is 15\arcsec and depends on the array direction
and the local time of observation.

The data from Nobeyama Radioheliograph \cite{Nakajima94}, Nobeyama
Radio Polarimeters (NoRP; \opencite{Nakajima85};
\opencite{Torii}), and Radio Solar Telescope Network (RSTN) have
been used to examine spatial and spectral characteristics of
background microwave bursts. The data from {\it Reuven Ramaty High
Energy Solar Spectroscopic Imager} (RHESSI) have been exploited to
analyze X-rays from the nonthermal plasma component \cite{Lin}.
Some events have been studied using UV images of flares acquired
by the {\it Transition Region and Coronal Explorer} (TRACE,
\opencite{Handy}). The magnetic field structure of flares was
inferred from the Michelson Doppler Imager (MDI) magnetograms
\cite{Scherrer}.

\section{Observations}
\begin{figure}    
  \centerline{\includegraphics[width=\textwidth]{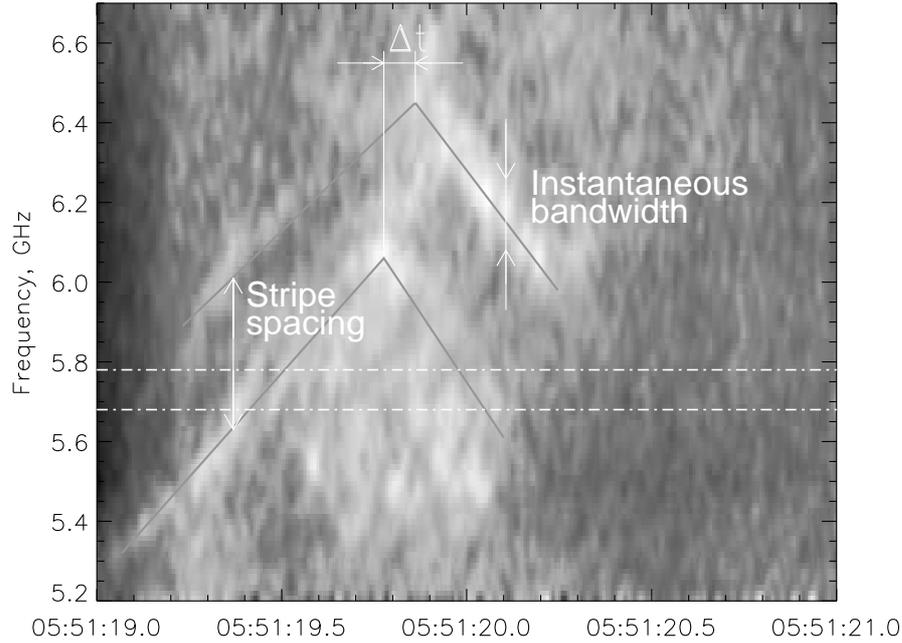}
              }
              \caption{The dynamic spectrum of the zebra structure observed
              on 17 September 2002. White means increased emission.
              The zebra stripes are indicated by the grey lines.
              The horizontal lines mark the boundaries of the SSRT receiver bands.
                      }
   \label{F1-simple}
   \end{figure}

Figure~\ref{F1-simple} presents the dynamic spectrum with a
simple, two-stripe zebra structure. This structure was recorded at
the Huairou station at the maximum of a short-duration microwave
burst from a 17 September 2002 solar flare. The small flare of
GOES X-ray class C2.0 occurred in AR 10114 (S12W42). The microwave
burst with the maximum intensity of 60 sfu at 4 GHz lasted for
less than one minute. Its spectrum shape was typical of the
gyro-synchrotron emission. The simulation of the gyro-synchrotron
spectrum (\opencite{Ramaty1969}; \opencite{Ramaty1994}) shows that
the magnetic field in the emission source was weaker than
200~gauss (G).

    \begin{figure}    
     \centerline{\includegraphics[width=\textwidth]{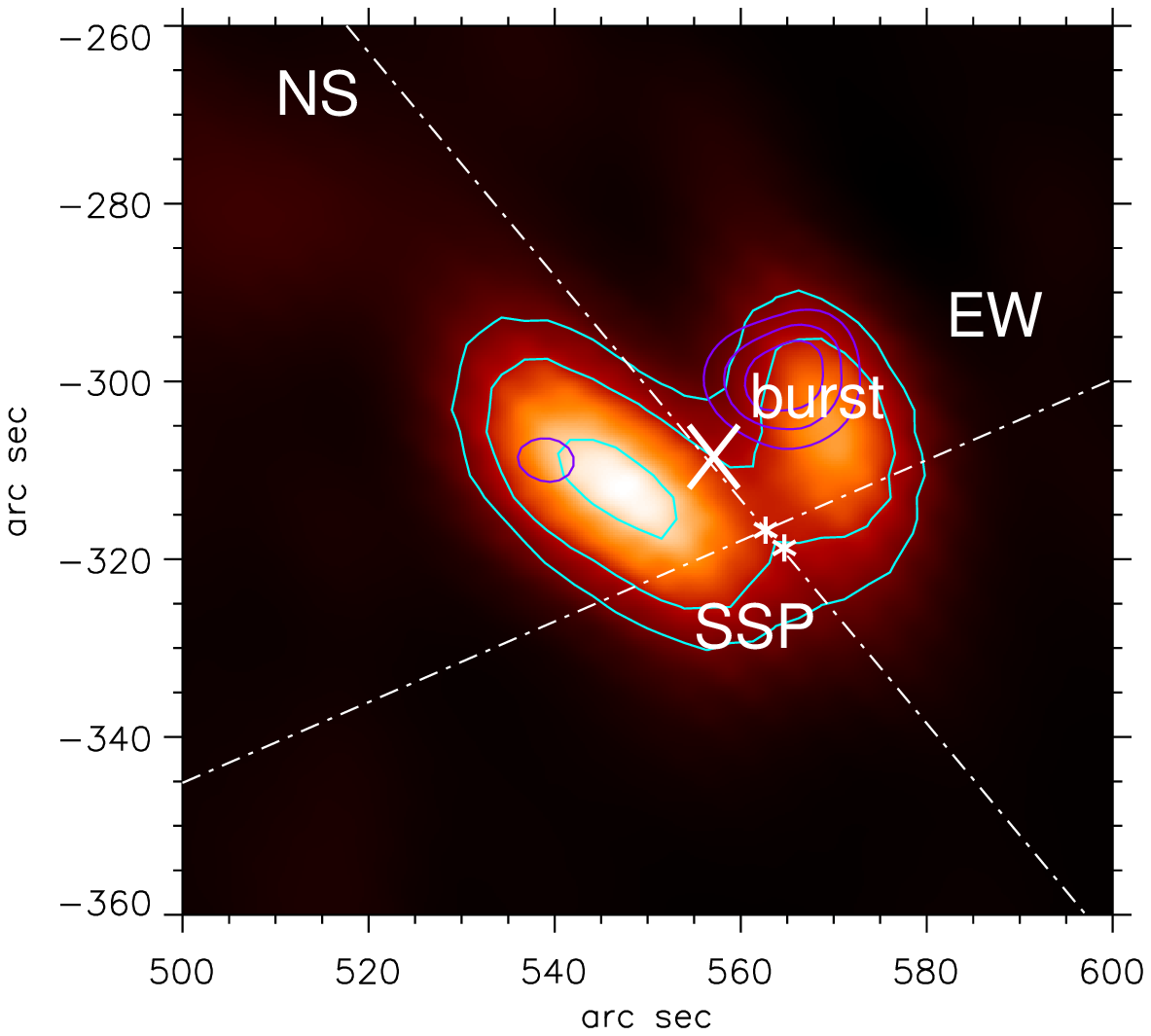}
              }
              \caption{The configuration of the 17 September 2002
              flare loop. The background image and blue contours
              at [0.3, 0.5, 0.9]$\times$0.2 MK show the brightness
              temperature at the frequency of 17 GHz (05:51:23 UT).
              The purple contours indicate X-ray counts in the energy
              range of 25 to 50 keV integrated from 05:50:40 to 05:51:40 UT.
              The dash-dotted lines show the scan directions of the SSRT
              linear interferometers. The big cross marks the center
              of the radio burst source at 5.7 GHz, and the asterisks
              indicate the sources of the zebra structures. The source
              of the descending branch in Figure 1 is more distant from
              the center of the burst.
                      }
   \label{F2-simple}
   \end{figure}

Narrow-band structures in the frequency range of 5.2-7.6 GHz were
observed near the burst maximum for 5 s from 05:51:19 to 05:51:24
UT. Among them, on the continuous background spectrum, we can
single out two zebras of about 1 s duration with a similar
structure and separated by 3 s. Figure~\ref{F1-simple} shows the
dynamic spectrum of the first structure. Two parallel stripes can
be identified which drifted to higher frequencies at a rate of
about 1~GHz~s$^{-1}$ initially, and later to lower frequencies at
a rate of 1.8~GHz~s$^{-1}$. The total bandwidth of this emission
slightly exceeded 1 GHz. The turnover in the drifting rate took
place in the high-frequency branch by $\Delta{t}$ = 85~ms later
than in the low-frequency one. The bandwidths of the emission
branches are 0.15 GHz (ascending branch) and 0.23~GHz (descending
one), and they are separated by 0.44 and 0.5~GHz, respectively.

During the flare, the interferometric observations at 17 GHz
showed a loop with bright footpoints separated by 25\arcsec
(Figure~\ref{F2-simple}). At the footpoints were compact X-ray
sources with emission at 25--50 keV. One-dimensional radio
brightness distributions recorded by the EW and NS linear arms of
SSRT at the frequency of 5.7~GHz indicated that the background
burst source covered the entire flare loop at 17~GHz. As
illustrated in Figure~\ref{F1-simple}, the lower-frequency stripe
of the zebra structure was emitted in the band of the SSRT
operating frequencies. The emission amplitude of the zebra stripes
did not exceed several solar flux units, and its sources in the
ascending and descending branches were located at the top of the
loop when the observation was made at 5.7~GHz. The relative
displacement between the sources of different branches was within
the accuracy of measurements of the centroid positions of radio
brightness and was no more than $3''$ in the plane of the sky.

\begin{figure}    
  \centerline{\includegraphics[width=\textwidth]{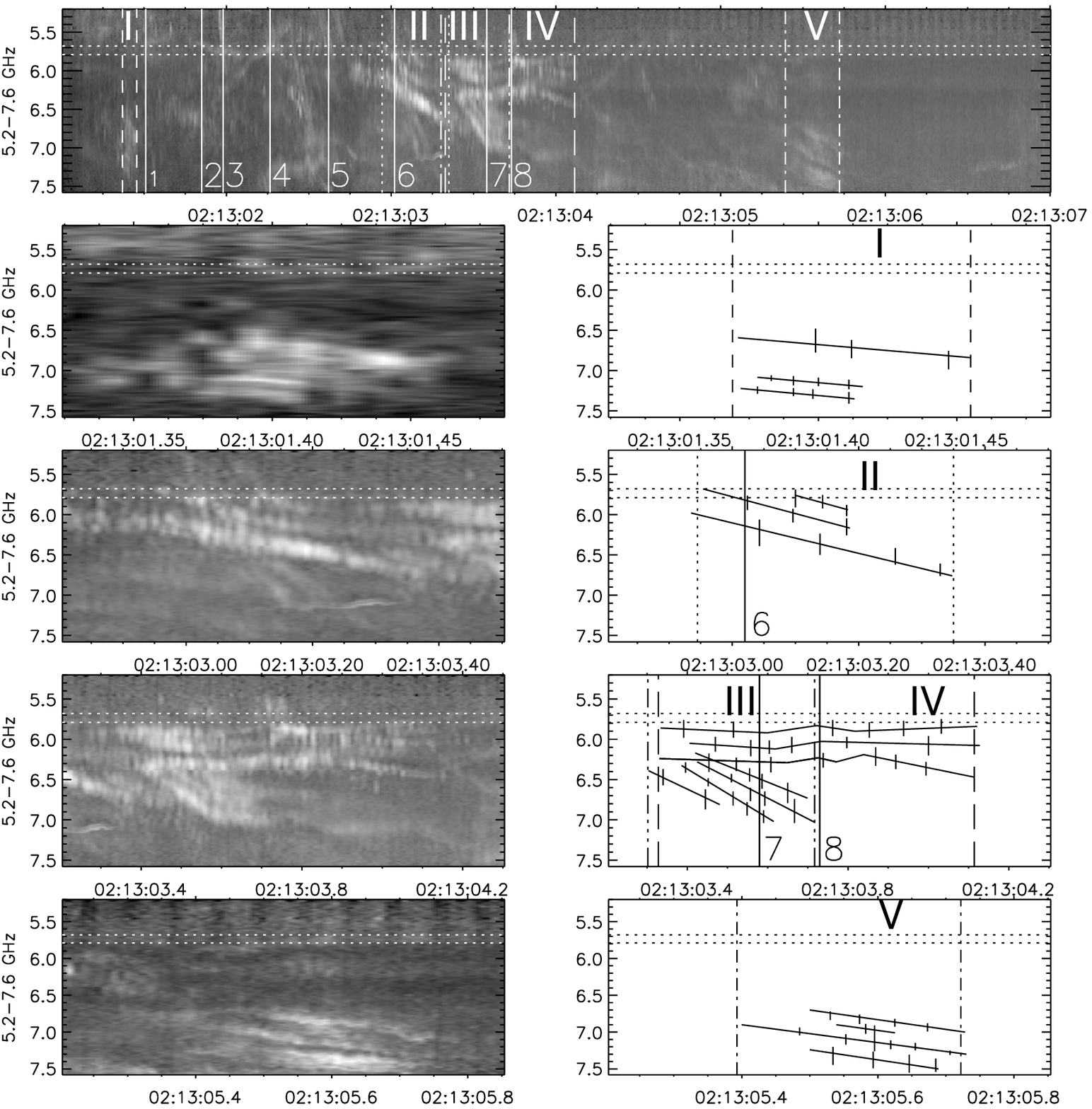}
              }
              \caption{The dynamic spectra with fine structures
              observed on 29 May 2003 (top panel).
              The four panels on the left show the enlarged
              portions marked by I - V. The panels on the right
              are schematic representations of zebra patterns
              seen in the corresponding left-hand-side panels.
              The individual zebra stripes are shown as solid
              lines with slopes representing the frequency drifts
              of individual stripes. The vertical bars show the
              instant bandwidths of drifting stripes in their
              bright portions. The two horizontal dashed lines
              indicate the frequency range of SSRT receivers.
              Arabic numerals attached to solid vertical lines
              mark the maximum moments of sub-second pulses recorded by the SSRT.
                      }
   \label{F3-simple}
   \end{figure}

A UV image of the flare at 195~\AA\ was available in a minute
after the zebra structure was recorded. The total area of the
flare brightening including the loop was about $4 \times 10^{17}
$cm$^2$. The GOES total flux observations show that the zebra
structures were observed during the ascending phase of the
emission measure at the constant temperature of $\approx$ 8 MK.
From the emission measure of $2 \times 10^{47} $cm$^{-3}$ follows
the plasma density of $\approx 7 \times 10^{10} $cm$^{-3}$, if we
assume the thickness of the emitting region to be $10^8$ cm.

Figure~\ref{F3-simple} presents the microwave dynamic spectrum
recorded during the 29 May 2003 solar flare \cite{Sych}. This
event was distinguished by the presence of several zebra
structures with three stripes (I, II, IV) and complex ones with
four stripes (III, V). The short burst with the main phase of only
$\approx$ 20 s (02:13:00 -- 02:13:20 UT) was associated with the
flare of GOES class M1.5/1F (02:09 -- 02:24~UT) in NOAA 10368
(S37E03). The zebra structures were observed at the maximum of the
background burst whose emission was as much as 180~sfu at 9.4~GHz.

The left-hand-side panels of Figure~\ref{F3-simple} show the zebra
structures expanded in time. The intensity of the stripes was
within several solar flux units. The stripes with various
frequency patterns and different drift rates were seen in a wide
frequency range (5.8--7.2 GHz) for 4 s. Pulses with sub-second
duration were recorded by the SSRT only during some structures: II
(6), III (7), IV (7, 8). Note that a very rare event was recorded
in which zebra stripes with different slopes were observed
simultaneously (III, IV). Probably, two independent sources (two
loops or two footpoints of the loop) were in the antenna beam.

The drifts were largely directed to higher frequencies. In the
second and fifth intervals, the frequency drift was $df/dt$ =
1.3--1.5~GHz~s$^{-1}$ at a frequency splitting of 200 to 300~MHz
between the stripes. In the first and third cases, $df/dt$ =
3.2--3.6~GHz~s$^{-1}$ at $\Delta{f}$ = 150--300~MHz. From
approximately 02:13:03.3 UT, two distinct and almost non-drifting
stripes separated by 300~MHz appeared between 5.8--6.4~GHz.

In this event the zebra patterns demonstrated the so-called
superfine temporal structure, namely the individual bright stripes
were consisted of clusters of separate short-duration pulses
\cite{Sych}. Their duration and bandwidth were around the
instrumental resolutions of 5 ms and 70-100 MHz, respectively.
Note that the superfine structure was also observed during two
other events on 10 April 2010 and 5 January 2003.

\begin{figure}
  \centerline{\includegraphics[width=0.8\textwidth,clip=]{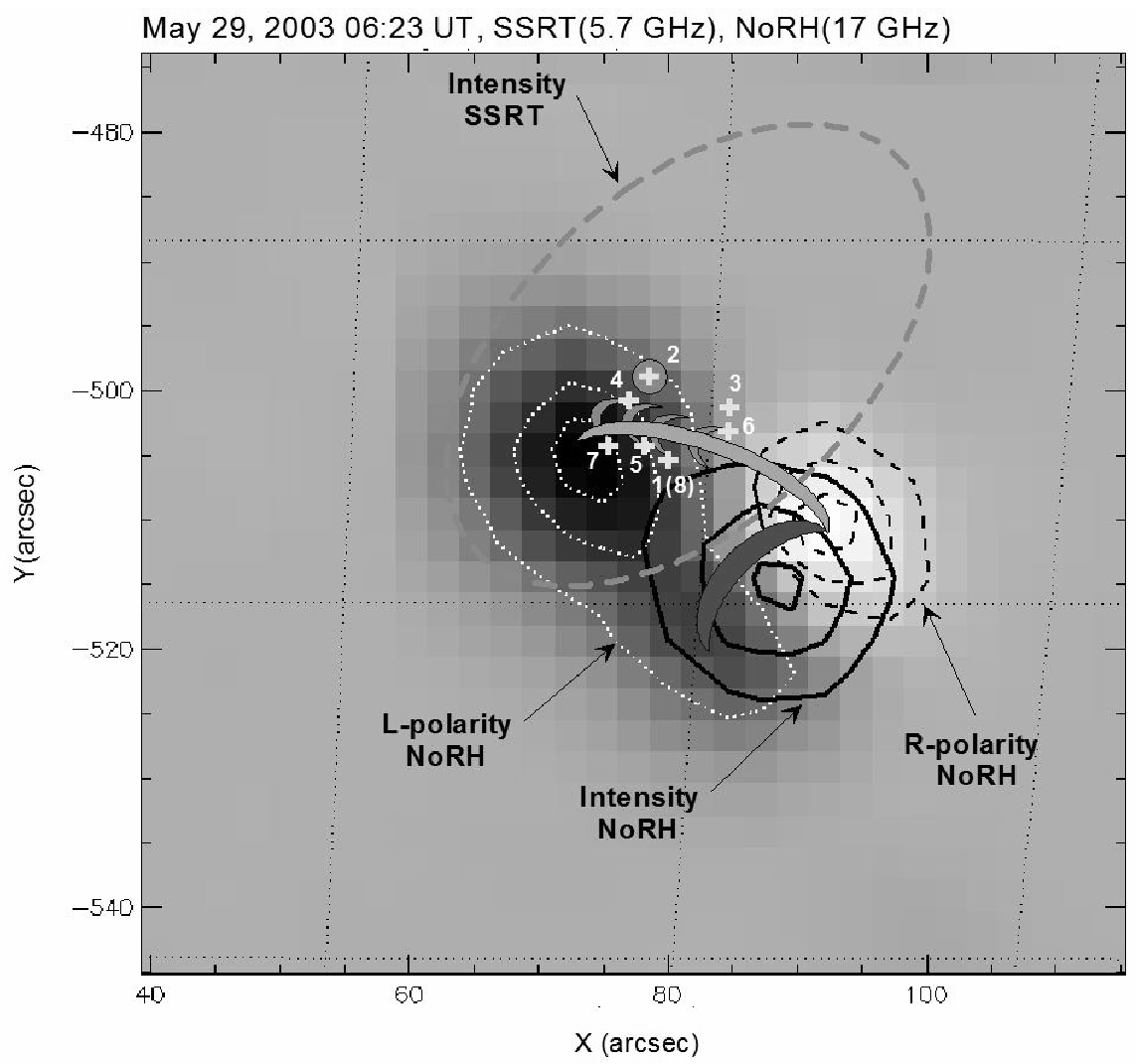}
                 }
              \caption{The positions of the radio sources at 5.7 GH
              (SSRT) and 17 GHz (NoRH) superimposed on the radio map
              at 17 GHz in polarization at 02:13:00 UT, 29 May 2003.
              The white region corresponds to right circular polarization
              at 17 GHz. Brightness of this source is shown by the dashed
              contours at 20, 50, and 80$\%$ of the maximum brightness (0.2 ~MK).
              The white dotted contours (20, 50, and 80$\%$ of the minimum
              brightness (-0.58 ~MK) show the structure of the background L-source.
              Solid black contours indicate 20, 50, and 80$\%$ of the maximum
              brightness (15.8 MK) of the 17 GHz source in intensity (R+L).
              The grey thick dashed contour marks 20$\%$ of the maximum brightness
              of the 5.7 GHz source (12 MK) at 02:17:54.8~UT. The pluses ($+$)
              indicate the positions of zebra sources 1--8.
                       }
   \label{F4-simple}
   \end{figure}

The configuration of the flare could be determined using sequences
of radio maps in intensity and polarization \cite{Sych}. During
the flare there were two polarized 17 GHz sources: a compact
source (right-handed, R) near the N-spot and a region extending
southwestward with opposite polarization (L)
(Figure~\ref{F4-simple}). The L-region exhibits two centers of
brightness on the northern and southern halves. The northern
source emerged near the S-polarity spot. The position of the
southern source corresponded to the S-polarity extension that was
elongated southward as the active region developed.

The flare was consisted of two loop systems joining the
R-polarization region and the two L-polarization centers. The
existence of a shorter loop in the south is consistent with the
peak position of the emission intensity at 17~GHz at the top of
the loop. Between the footpoints of the longer and higher loop was
located the region of flare emission at 5.7~GHz, which was
extended around the photospheric magnetic neutral line. The zebra
sources were in the coronal part of this loop system. The apparent
sizes of the sources were close to the SSRT beam size; therefore,
their true sizes were much less than 15\arcsec. The degree of
polarization of radio pulses was low and showed changing signs.
The significant difference among individual zebra structures
suggests that their sources were located in different magnetic
loops.

All the events with zebra structures detected in the dynamic
spectra in the frequency range of 4.5--7.5~GHz are listed in
Table~\ref{T-complex}. In all the cases, the frequency range of
the zebra structures was less than the dynamic range of the
spectro-polarimeter. We selected those events with at least two
simultaneous bright stripes varying identically with time. The
absorption bands (between bright stripes) with the brightness
temperature lower than the background one, which had been detected
occasionally at lower frequencies, were not observed in these
events (\opencite{Slottje}). The magnetic field strengths in the
zebra-structure sources were estimated by using the potential
field approximation, and are given for a number of events. For the
zebra stripes recorded with SSRT the polarization degree was
determined from the SSRT data; for other cases a similar procedure
was applied to the NOAC data.

\begin{kaprotate}
\begin{table}
\caption{ Parameters of zebra-patterns in six events. }
\label{T-complex}
\begin{tabular*}{\maxfloatwidth}{lcccccccccccccc}
  \hline
Event &  \multicolumn{3}{c}{10 Apr 01\footnotemark[1] }& 21 Aug 02
&
      \multicolumn{2}{c}{17 Sep 02} & \multicolumn{2}{c}{05 Jan 03\footnotemark[2]} & 18 Mar 03\footnotemark[3] & \multicolumn{5}{c}{29 May 03\footnotemark[4]} \\
      \hline
Number of   & 4  &  2 & 2 & 2 & 2 & 2 & 4 & 4 & 4 & 3 & 3& 3& 2&3 \\
stripes     &    &    &   &   &   &   &   &   &   &   &  &  &  &  \\
      \hline
$\Delta {f}$ & 0.09-- & 0.11& 0.12--&0.63--&0.44--&0.20--&0.16&0.22&0.3--&0.14--&0.23--&0.14--&0.13--&0.14--\\
       $[$GHz$]$   & 0.2  &     & 0.24&0.73&0.50&0.38&    &    &0.5&0.48&0.27&0.27&0.29&0.21\\
      \hline
Durations&1.2&0.8 &2.1&0.9&1.3&0.4&2.4&4.1&0.4&0.09&0.45&0.3&0.8 &0.34\\
$[$s$]$        &   &    &   &   &   &   &   &   &   &    &    &   &    & \\
      \hline
Instant      &0.1&0.04& 0.35&0.26&0.2&0.09&0.06 &0.26&0.12--&0.04--&0.1--&0.16--&0.12--&0.1--\\
stripe width,&   &    &     &    &   &    &     &    &0.29&0.2 &0.16&0.32&0.16&0.17\\
$[$GHz$]$          \\
      \hline
Drift rate,  &-1.0&-1.4&-0.3&+2.5&-1.0&+2.7&-0.7&   &-11. &+3.4&+2.1&+2.85&+0.39&+1.27\\
$[$GHz~s$^{-1}$$]$        &+0.2&+0.6&+1.0&    &+1.8&    &-0.5&   &+19.6&+4.9&+2.7&+3.23&+0.44&+1.81\\
      \hline
Delay     &$<$40&$<$10&$<$10&50& 85&    &$<$10&  &10--\\
      $[$ms$]$                   &     &     &     &  &   &    &     &  &130\\
      \hline
Polarization&30--&30--&&5--&&0&100&100&20&5&90&50&50&50\\
$[$\%$]$           &50  &50  &&30\\
  \hline
\end{tabular*}
$^1$ Chernov {\it et al.} (2006)

$^2$ Altyntsev {\it et al.} (2006)

$^3$ Ning {\it et al.} (2007)

$^4$ Sych {\it et al.} (2010)

\end{table}
\end{kaprotate}

Among the sources presented in Table~\ref{T-complex}, the 5
January 2003 zebra structure stands out. This event exhibited four
stripes separated uniformly in frequency  and synchronously
changing frequency drifts. The event was studied in detail by
\inlinecite{Altyntsev2005}. A frequency gap between adjacent
stripes corresponded to the cyclotron frequency in the source
region. The emission from the compact zebra source was X-mode
polarized up to 100$\%$ degree.

The stripes observed in the 10 April 2001 event were equally
spaced in frequency  within the measurement accuracy. The analysis
of this event \cite{Chernov2006b} has demonstrated, however, that
the emission mechanism with the Bernstein modes is inapplicable in
this event because the frequency gaps between the adjacent stripes
were too small. In the other events, the frequency separation
between the stripes was not uniform. Usually the observations
showed no more than four stripes. In some cases, a time delay was
seen between the two similar stripes. The characteristic
parameters of zebra structures are summarized in
Table~\ref{T-simple}.

\begin{table}
\caption{The observed characteristics of zebra structures
}
\label{T-simple}
\begin{tabular}{ll}
\hline
Frequency range $f$ [GHz] & 5.2 -- 7.5\\
Number of stripes & 2 -- 4 \\
Frequency gaps between stripes $\Delta {f}$ [GHz] & 0.1 -- 0.7 \\
Duration [s] &  0.1 -- 4.1\\
Instant frequency width of a stripe [GHz] & 0.04 -- 0.35\\
Drift rate [GHz~s$^{-1}$]&  0.4 -- 19\\
Polarization [$\%$]& 0 -- 50\\
Delay $\Delta {t}$ [ms]  & $<$ 130 \\
\hline
\end{tabular}
\end{table}

\section{Discussion and Conclusions}

The observations of zebra structures with high spatial resolution
have revealed that their sources are located in the coronal part
of flare loops in which the magnetic field does not exceed several
hundred G. The narrow frequency bandwidth of zebra-structure
stripes suggests a coherent mechanism for their generation. The
coherent mechanism responsible for the continuous background
emission will be characterized by the Langmuir frequency $f_{\rm
p}$ determined by the plasma density $n$, and by the electron
cyclotron frequency $f_B$ which depends on the magnetic field
strength $B$. The Langmuir frequency exceeds 4 GHz at a plasma
density of $2 \times 10^{11} $cm$^{-3}$. If the electromagnetic
wave is emitted at twice the Langmuir frequency, the density in
the source is about $5 \times 10^{10} $cm$^{-3}$. The cyclotron
frequency reaches this value in the magnetic field over 1400~G
(characteristic for flare loop footpoints in the chromosphere)
while in the corona the magnetic field is much weaker. Hence, in
the zebra sources the condition $f_{\rm p} \gg f_B$ can be
considered fulfilled.

In the events with simple configuration (17 September 2002 and 5
January 2001) with one flare loop, the GOES soft X-ray
observations enabled us to estimate the plasma density in the
source region. We found that the zebra frequencies were about
twice the Langmuir frequency there. The coherent emission is
caused by nonthermal electrons with pitch-angle anisotropy which
will relax into equilibrium through excitation of electrostatic
oscillations due to the development of loss-cone instability. In
flare loops the electron beams with transverse anisotropy in the
velocity distribution were observed by \citeauthor{Altyntsev2007}
(\citeyear{Altyntsev2007}, \citeyear{Altyntsev2008}).

The first microwave zebra was found in the dynamic spectra of the
flare of 5 January 2003 and described in
\inlinecite{Altyntsev2005}. The frequency gap between adjacent
bright stripes was constant and close to the cyclotron frequency.
The SSRT observations revealed that there is only one emission
source for multiple zebra stripes in the compact region. The
position of the source relative to the magnetic field and the
sense of polarization suggests the extraordinary mode of emission.
The plasma density in the zebra source was estimated from soft
X-ray data. The knowledge of plasma parameters in the source
allowed us to identify the observed emission as the excitation of
the Bernstein modes of magnetized plasma oscillations, namely the
field-aligned electrostatic waves at harmonic frequencies of the
cyclotron frequency (\opencite{Altyntsev2005};
\opencite{Kuznetsov2005}).

In other events with three or four stripes, the zebra stripes were
not equally-spaced in frequency. In this case, the most probable
mechanism for generating zebra emission is the so-called double
plasma resonance (DPR) (\opencite{Zheleznyakov};
\opencite{Zlotnik2009}). These individual stripes are assumed to
be generated in different flare sites near the so-called resonance
surfaces. On these surfaces the condition $f^s=\sqrt{f_{\rm p}
(n^s)^2+f_B (B^s)^2)}=s f_ B (B^s)$ should be met, where $n^s$ is
the plasma density and $B^s$ is the magnetic field strength on the
resonance surface corresponding to the $s$-th harmonic. Under such
condition the growth rate $\gamma$ of the kinetic instability of
upper hybrid waves is large, $\gamma$ $\sim \frac{n_{\rm ac}}{n}f_
B$, and is much higher than the growth rate of the Bernstein-mode
instability, where $n_{\rm ac}$ is the density of accelerated
electrons.

The analysis of zebra characteristics in a meter-wave range has
supported the DPR emission model (\opencite{Aurass};
\opencite{Zlotnik2003}). The 25 October 1994 event in AR 7792 with
two interacting loops with scales differing by an order of
magnitude has been examined. The magnetograms of this active
region were used to calculate the magnetic field distribution
along a large coronal loop. These calculations and zebra frequency
structure have led to the plasma density distribution along the
loop, which proved to be close to the expected hydrostatic
distribution at a reasonable value of coronal plasma temperature
of one million degrees.

The frequency gap in emission from different resonance surfaces
should be equal: $\Delta{f}=2f^{(s+1)}-2f^s$. If $f_{\rm p} \gg
f_{\rm B}$, the plasma density and the magnetic field strength in
the sources of adjacent stripes are related as follows:

\begin{eqnarray}
\frac{B^{(s+1)}}{B^s}
=\frac{s}{s+1}\left(1+\frac{\Delta{f}}{2f^s}\right)
\end{eqnarray}

\begin{eqnarray}
\frac{n^{(s+1)}}{n^s} =\left(1+\frac{\Delta{f}}{2f^s}\right)^2
\end{eqnarray}

In the 17 September 2002 event, the estimated magnetic field
strength is 200~G and the corresponding cyclotron frequency is
0.56~GHz. The frequency of the zebra structure may be as high as
6.4~GHz. If the transformation into electromagnetic waves occurs
at twice the cyclotron frequency, then we obtain $s \approx$ 5.
Given the difference in frequencies between stripes $\Delta{f}$ =
0.44~GHz $\leq f_B$, we obtain the relations
$\frac{B^{(s+1)}}{B^s} \approx$ 0.94 and $\frac {n^{(s+1)}}{n^s}
\approx 1.07$ between the magnetic field strength and plasma
density in the sources of the stripes. It is evident that, in the
case of small harmonic numbers, the changes in the magnetic field
strength and density in the sources of adjacent stripes have
opposite signs.

In the DPR model, the resonance conditions are determined by the
magnetic field strength and plasma density, which can change with
magnetohydrodynamic velocities and lead to the frequency drift of
zebra stripes. The appearance and lifetime of zebra structures
depend on the presence of nonthermal electrons with pitch-angle
anisotropy in the resonance regions. A theoretical study on the
excitation of zebra structure  shows their critical dependence on
the velocity and angular distributions of nonthermal electrons
(see \opencite{Zlotnik2009}). The mechanism of how DPR is
generated under the condition of parameters close to the microwave
observations with SSRT was examined in detail by
\citeauthor{Kuznetsov2005} (\citeyear{Kuznetsov2005},
\citeyear{Kuznetsov2007a}, \citeyear{Kuznetsov2008}) and
\inlinecite{Kuznetsov2007b}. Given a plasma temperature of about
ten MK in a flare, the growth rate of loss-cone instability
exceeds the Coulomb collision time if the nonthermal electron
density satisfies $\frac{n_{\rm ac}}{n}>10^{-6}$. The theoretical
analysis suggests that the intensity and the polarization degree
of the emission may vary in a wide range and take the maximal
values in a narrow angular range perpendicular to the magnetic
field. The strong angular dependence of emission intensity is one
of the reasons why they are so rarely observed.

 \begin{figure}    
   \centerline{\includegraphics[width=0.8\textwidth,clip=]{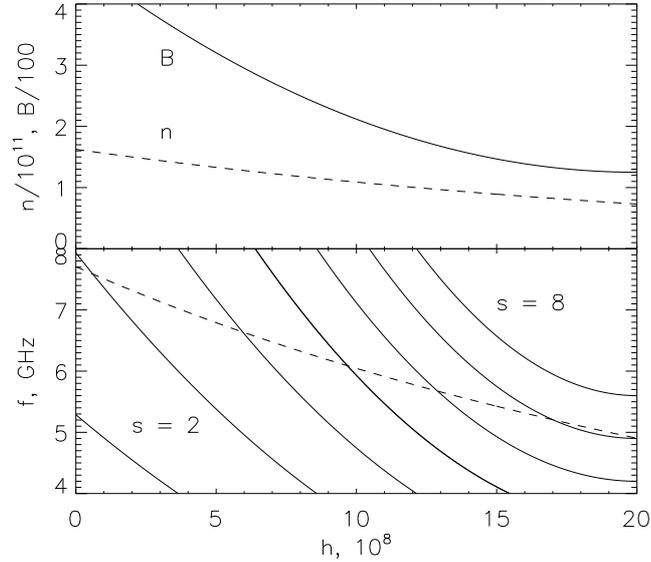}
              }
              \caption{Top: Plasma density and magnetic field
              strength as a function of height. Bottom: Harmonics
              of the doubled cyclotron frequency (solid lines) and
              the upper-hybrid frequency (dashed line). The fifth
              cyclotron harmonic is shown by the thick solid curve.}
   \label{F5-simple}
   \end{figure}

Zebra structures were observed both in the disk center and near
the limb. Therefore, the position of the source is supposed to be
near the top of the flare loop where the density gradient along
the loop is small. However, the DPR model with radio sources
distributed along the loop requires higher density gradient as
compared to the hydrostatic value \cite{Chernov2006b}. Figure
\ref{F5-simple} gives an example of how the location of the
resonance surface is calculated. The calculations were made on the
assumption that the frequency is doubled during the transformation
of electrostatic oscillations into electromagnetic waves. The
profile of the magnetic field along the loop is described by
$B(h)=B_{\rm top}\left(1+\left(\frac{h}{L_B}\right)^2\right)$
\cite{Lee}. The graph for $B(h)$ in the upper panel of Figure
\ref{F5-simple} was derived by assuming $B_{\rm top}$= 125~G and
$L_B$ = 12~Mm which are typical for flare loops. In the lower
panel of Figure \ref{F5-simple}, the solid lines indicate the
doubled frequencies of cyclotron harmonics in the range $s = 2-8$.
The dashed lines show the plasma density (upper panel) and the
doubled upper-hybrid frequency (lower panel) vs. height. The
intersections of the dashed line with the solid curves in the
lower panel of Figure \ref{F5-simple} correspond to the resonance
frequencies and their positions along the flare loop. The
intersection of the upper-hybrid frequency (derived from the
assumption of hydrostatic density distribution) and the fifth
harmonic of the cyclotron frequency gives the frequency of
6.4~GHz, and the frequency gaps $\Delta f$ are comparable to the
observed ones as has been assumed. The simulation, however, shows
that the plasma temperature ($\leq$ 0.5 MK) in the loop should be
several times lower than that of an ordinary coronal loop and by
an order of magnitude lower than those characteristic of the flare
plasma.

\begin{figure} 
\centerline{\includegraphics[width=0.8\textwidth,clip=]{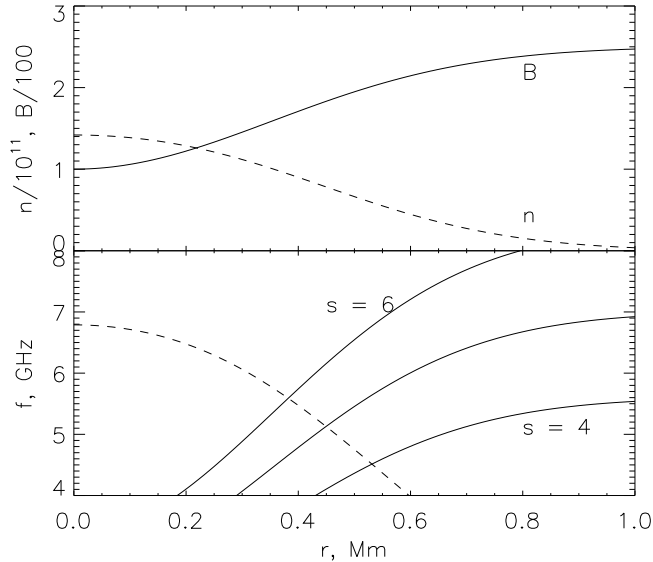}
}
\caption{%
Similar to Figure 5 but for an equilibrium magnetic flux tube. }
\label{F6-simple}
\end{figure}

The frequency drift of brightness stripes in zebra structures
reflects changes in plasma parameters in the source region. In
order to retain the resonance, the equality between the Langmuir
frequency and the cyclotron frequency harmonic requires a
correlated temporal changes in plasma and magnetic field
parameters as $n \sim B^2$. Such a relation between density and
magnetic field is typical of the transverse equilibrium of a
current rope. Let us consider the simplest case of a static
magnetic tube in which the radial equilibrium condition
$\frac{B^2(r)}{8\pi}+2k_{\rm B} n(r)T=\frac{B_{\rm ext}^2}{8\pi}$
is met, where $k_{\rm B}$ is the Boltzmann constant and $B_{\rm
ext}$ is the external magnetic field. It is assumed that plasma
pressure can be ignored outside the loop. Figure \ref{F6-simple}
presents the calculation results for this model analogous to those
in Figure \ref{F5-simple}. The plasma temperature inside the tube
is taken to be constant and equal to 8 MK. The radial magnetic
field distribution is assumed to be $B$=250--150
$\exp{\left(-r^2/p^2\right)}$ [G], and the plasma density
distribution is determined from the equilibrium equation. The
scale $p$ in the magnetic field is a free parameter and it is
taken as 0.5~Mm.

\begin{figure} 
\centerline{\includegraphics[width=0.8\textwidth,clip=]{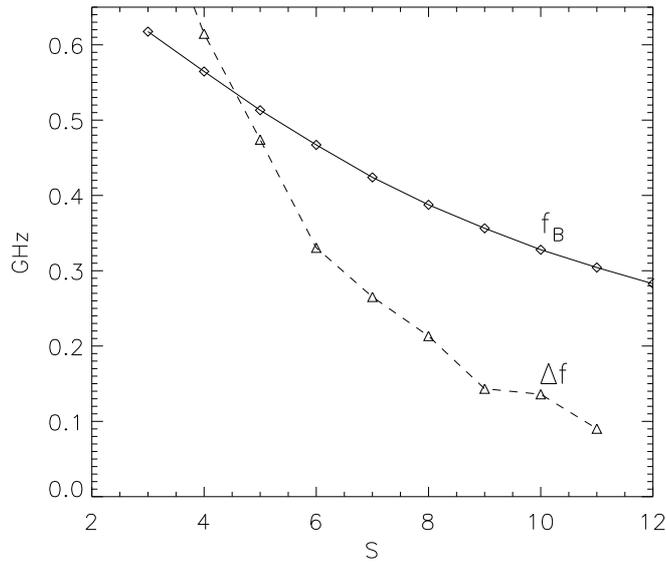}
}
\caption{%
The calculated frequency gaps between adjacent strips (dashed
line) in comparison with the cyclotron frequency on the
corresponding surfaces (solid curve). } \label{F7-simple}
\end{figure}

Thus, a simple model with reasonable assumptions about plasma
parameters can satisfactorily describe the observed frequency gaps
between zebra stripes (Figure \ref{F7-simple}). The frequency gaps
will be higher than the cyclotron frequency at small harmonic
numbers and lower at large harmonic numbers. The model has a
number of free parameters such as the radial profile of magnetic
field and plasma temperature. By choosing these parameters, we can
obtain agreement between the calculated and observed zebra
structures. It seems reasonable to attribute the frequency drift
of zebra stripes to sausage oscillations which generate an
increase of plasma density and magnetic field in compressible
magnetic tubes. The observed frequency range of about 1 GHz is
achieved for a relatively small change in plasma density ($\approx
30\%$).

In order to explain the narrow frequency band of a single strip,
the longitudinal size of the emitting region should be much less
than the characteristic scales of plasma density and magnetic
field variations. The upper bound on the size of the emitting
region can be estimated considering the absorption of
electromagnetic waves in the plasma surrounding the magnetic tube.
According to \inlinecite{Benz} the optical thickness at twice the
Langmuir frequency is $\tau=1.2{T_{\rm keV}}^{-3/2}\nu_{\rm GHz}^2
\lambda_{\rm cm}$, where $\lambda$ is the radial scale of density
variations. At $\lambda = 0.5$ Mm, we obtain $\tau \approx 0.1$,
and the absorption when the emission escapes across the magnetic
tube is negligible. However, the absorption may be substantial if
the loop radius is more than several thousand kilometers and the
emission is observed in the direction along the loop.

The superfine structures observed in the three zebra events are
not the main subject of this study. Observations of such
structures were described by \inlinecite{Chernov1998};
\inlinecite{Chernov2003}; \inlinecite{Meszarosova}.
\inlinecite{Kuznetsov2008} and \inlinecite{Rozhansky} have
concluded that superfine temporal structures are formed due to
modulation of the emission mechanism by MHD turbulence in the
zebra sources. To promote understanding of the phenomena we will
be able to exclude the propagation effects by conducting
simultaneous observations using multiple radio telescopes. There
is only 10 April 2001 event which was observed simultaneously with
the PMO and the Huairou station. In Figure 2 of
\inlinecite{Chernov2006b} all smooth changes in the zebra stripes
were strictly identically reproduced in the spectra from both
observatories, but there was no detailed correspondence between
the fine structures of the stripes.

Up-to-date high-sensitivity spectro-polarimeters enabled us to
record several events with microwave zebra structures. Among two
hundred events with fine time structures recorded by Chinese
spectro-polarimeters and SSRT in the frequency range from 5.2 to
7.5 GHz, only six events had zebra structures. Their intensity did
not exceed several solar flux units. The observed stripes in the
microwave range corresponded to small numbers of harmonics-- near
the fifth--, and there were no more than four stripes
simultaneously. The frequency gaps between adjacent stripes varied
over a wide range from 0.1 to 0.7~GHz. The characteristics of the
observed events are consistent with our knowledge about the
emission mechanism (the double plasma resonance) and can be
applied to the diagnostics of plasma conditions in coronal flare
loops.

\section*{Acknowlegement}
We appreciate discussions with Dr. E.Ya. Zlotnik.  The research
carried out by Robert Sych at NAOC was supported by the Chinese
Academy of Sciences Visiting Professorship for Senior
International Scientists, grant No. 2010T2J24. This study was
supported by the Russian Foundation of Basic Research
(08-02-00270, 08-02-92204-GFEN, 09-02-92610, 09-02-00226,
10-02-00153a). This research was supported by a Marie Curie
International Research Staff Exchange Scheme Fellowship within the
7th European Community Framework Programme.

\end{article}
\end{document}